\documentclass[numbers]{aipproc}
\layoutstyle{8x11double}
\usepackage{graphicx}
\usepackage{amsmath}
\usepackage{amsfonts}
\usepackage{amssymb}
\usepackage{feynmf}
\usepackage{hyperref}

\allowdisplaybreaks

\def\siml{{\ \lower-1.2pt\vbox{\hbox{\rlap{$<$}\lower6pt\vbox{\hbox{$\sim$}}}}\ }}
\def\simg{{\ \lower-1.2pt\vbox{\hbox{\rlap{$>$}\lower6pt\vbox{\hbox{$\sim$}}}}\ }}

\def \br {\mathbf{r}}

\def \bx {\mathbf{x}}
\def \by {\mathbf{y}}

\def \crr {C_R}
\def \nc {N}
\def \ca {C_A}
\def \nf {n_f}

\def \trt {\tilde{\mathrm{Tr}}}
\def \als {\alpha_{\mathrm{s}}}

\def \m2   {\mu^{2 \epsilon}}

\begin{document}
\title{The Polyakov loop correlator at NNLO and singlet and octet correlators}

\author{Jacopo Ghiglieri}{
address={Physik-Department, Technische Universit\"at M\"unchen,
James-Franck-Str. 1, 85748 Garching, Germany},
altaddress={Excellence Cluster Universe, Technische Universit\"at M\"unchen, 
Boltzmannstr. 2, 85748, Garching, Germany}}

\begin{abstract}
We present the complete next-to-next-to-leading-order calculation of the correlation function of two Polyakov loops for temperatures smaller than the inverse distance between the loops and larger than the Coulomb potential. We discuss the relationship of this correlator with the singlet and octet potentials which we obtain in an Effective Field Theory framework based on finite-temperature potential Non-Relativistic QCD, showing that the Polyakov loop correlator can be re-expressed, at the leading order in a multipole expansion, as a sum of singlet and octet contributions. We also revisit the calculation of the expectation value of the Polyakov loop at next-to-next-to-leading order.
\end{abstract}

\date{\today}

\maketitle

\section{Introduction}
The Polyakov loop and the correlator of two Polyakov loops are the order parameters of the deconfinement transition in pure gauge $SU(N)$ theory \cite{Kuti:1980gh,McLerran:1981pb}. As such, they have been extensively measured on the lattice, both in pure gauge and in full QCD (for a review see \cite{Petreczky:2005bd}). From a phenomenological point of view, the Polyakov loop and the correlator are important because they are related respectively to the free energy of a static quark and of a static quark-antiquark pair in a thermal bath. As such, they can be used as input for the phenomenology of heavy quarkonia in the quark-gluon plasma through potential models (see for instance \cite{Mocsy:2007jz}). However, although the spectral decomposition of the correlator is known \cite{Jahn:2004qr,Luscher:2002qv}, its relationship to the in-medium heavy quark potential is still in need of clarifications \cite{Philipsen:2008qx}, especially after the recent developments in a derivation of the potential from QCD \cite{Laine:2006ns,Beraudo:2007ky} and its rigorous definition and calculation within an Effective Field Theory (EFT) framework \cite{Brambilla:2008cx,Escobedo:2008sy,Brambilla:2010vq}.\\
The work of Ref. \cite{plc}, which we shortly present here, aims at a better understanding of the Polyakov loop correlator in the short distance regime $rT\ll 1$. This is achieved by performing first a perturbative NNLO (order $g^6$) computation of the correlator, improving the leading-order results of \cite{McLerran:1981pb,Gross:1980br} and complementing the order $g^6$ results of Nadkarni \cite{Nadkarni:1986cz}, which are valid for $rm_D\sim 1$, $m_D$ being the Debye screening mass. As a side result, we also obtain the perturbative expression of the Polyakov loop at the NNLO (order $g^4$). 
We then proceed to construct an EFT framework, based on potential Non-Relativistic QCD (pNRQCD), which allows us to reobtain the previous result for the correlator in terms of the colour-singlet and colour-octet zero-temperature potentials of pNRQCD and further contributions, which we compute, induced by the thermodynamical scales.
\section{Perturbative results for the Polyakov loop and its correlator}
We briefly present here the calculation methodology and the results for the Polyakov loop and its correlator. The former is defined as
$\langle L_R\rangle=\langle\trt L_R\rangle$, where  $L_R\equiv\mathrm{P}\exp\left(ig\int_0^{1/T}d\tau A^0(\bx,\tau)\right)$ is the Polyakov line in a colour representation $R$, either the fundamental $F$ or the adjoint $A$, and $\trt$ is the colour trace normalized by the dimension of the representation $R$. The correlator is defined as $\langle\trt L^\dagger_F({\bf 0})\trt L_F(\br)\rangle$; what we consider in the following is its connected part $C_\mathrm{PL}(r)$ in the fundamental representation.\\ 
We adopt the static gauge \cite{static}, defined by the condition $\partial^0A^0=0$. In this gauge the Polyakov line becomes local in time, i.e. $L=\exp\left((igA^0(\bx))/T\right)$. The one-loop expression of the 00 component of the gluon self-energy in this gauge, which is a fundamental ingredient in both computations, was not fully known in the literature and has been computed in \cite{plc}.\\
Both calculations assume a weak-coupling hierarchy $T\gg m_D$, where $m_D^2=g^2T^2/3(\nc+\nf/2)$, with $n_f$ light quarks. In the case of the correlator we furthermore assume $1/r\gg T\gg m_D\gg g^2/r$, where the last scale is the Coulomb scale that appears naturally in a perturbative quark-antiquark correlator. We stop our perturbative calculation at order $g^6(rT)^0$, meaning that we do not show terms with powers of $g$ greater than 6, and at order $g^6$ we do not show powers of $rT$ greater than zero. The aforementioned hierarchies are implemented by separating the momentum regions in the integrations by appropriate expansions, and by resumming the Debye mass in the longitudinal propagators when the loop momentum is of the same order. All divergences are treated in dimensional regularization; the Polyakov loop is finite at the order considered, whereas the correlator requires the usual charge renormalization of QCD.\\
Our result for the Polyakov loop is
\begin{eqnarray}
\nonumber \langle L_R\rangle&=&1+\frac{\crr \als }{2}\frac{m_D}{T}+\frac{\crr\alpha^2_s}{2}
\left[\ca\left(\ln\frac{m_D^2}{T^2}+\frac{1}{2}\right)\right.\\
&&\hspace{2.5cm}-n_f\ln2\bigg]+\mathcal{O}(g^5).
\label{finalg4loop}
\end{eqnarray}
Our result differs from the long-time accepted result of Gava and Jengo \cite{Gava:1981qd} in the finite terms at order $g^4$ and agrees with another recent determination \cite{Burnier:2009bk}. For the correlator we have
\begin{eqnarray}
&&C_{\mathrm{PL}}(r,T)=
\frac{\nc^2-1}{8\nc^2}\left\{ 
\frac{\als(1/r)^2}{(rT)^2}
-2\frac{\als^2}{rT}\frac{m_D}{T} \right.
\nonumber\\
&&
\hspace{0.8cm}
+\frac{\als^3}{(rT)^3}\frac{\nc^2-2}{6\nc}
+ \frac{1}{2\pi }\frac{\als^3}{(rT)^2}\left(\frac{31}{9}\ca-\frac{10}{9}n_f \right.
\nonumber\\
&&\hspace{0.8cm}+2\gamma_E\beta_0\bigg)
+ \frac{\als^3}{rT}\left[
\ca\left(-2 \ln\frac{m_D^2}{T^2} + 2-\frac{\pi^2}{4}\right)\right. 
\nonumber\\
&&
\hspace{0.8cm}
\nonumber\left.+ 2n_f\ln 2\bigg]
+\als^2\frac{m_D^2}{T^2} -\frac{2}{9}\pi \als^3 \ca
\right\}\\
&&
\hspace{0.8cm}
+\mathcal{O}\left(g^6(rT),\frac{g^7}{(rT)^2}\right),
\label{finalcpltot}
\end{eqnarray}
where the scale of the running coupling in the first, leading-order term is fixed to $1/r$ by the higher-order terms. Nadkarni \cite{Nadkarni:1986cz} computed the same quantity under the assumption $T\gg1/r\sim m_D$; as such, a direct comparison between the two computations is not possible. However the part of our results that does not depend on the hierarchy $rT\ll1$ agrees with Nadkarni's results expanded for $rm_D\ll1$.  
\section{The Polyakov-loop correlator in an EFT language\label{sec_EFT}}
We now proceed to briefly illustrate the principles and the construction of the EFT framework of \cite{plc}. Such a framework allows us to reobtain the result of Eq. \eqref{finalcpltot} in terms of colour singlet and colour octet correlators, making more explicit the physical meaning of the calculation.\\
The basic principle is that we implement the hierarchy $1/r\gg T\gg m_D\gg g^2/r$ at the Lagrangian level, by integrating out from (static) QCD these scales. We start from the largest, the inverse distance $1/r$. This leads to pNRQCD. Since $1/r$ is larger than all other scales, thermal ones included, those can be put to zero in the matching and the Lagrangian is the usual zero-temperature one \cite{pnrqcd}. In Euclidean space-time the action reads
\begin{eqnarray}
S_{\rm pNRQCD} &=&  
\int_0^{1/T} \!\! d\tau \int_\bx \int_\br\, {\rm Tr} \Bigg\{ {\rm S}^\dagger (\partial_0+V_s){\rm S}
+ {\rm O}^\dagger (D_0
\nonumber\\
&& 
+V_o){\rm O}- iV_A \left( {\rm S}^\dagger {\bf r}\cdot g{\bf E} {\rm O} + {\rm O}^\dagger {\bf r}\cdot g{\bf E} {\rm S}\right)
\nonumber\\
&&  
- \frac{i}{2}V_B \left({\rm O}^\dagger {\bf r}\cdot g{\bf E} {\rm O} + {\rm O}^\dagger{\rm O} {\bf r}\cdot g{\bf E} \right) + \frac{i}{8} V_C 
\nonumber\\
&&
\times\left(r^i r^j  {\rm O}^\dagger D^igE^j {\rm O} - r^i r^j  {\rm O}^\dagger {\rm O} D^igE^j \right)\nonumber\\
&&+ \delta {\cal L}_{\rm pNRQCD}
\Bigg\}+ \int_0^{1/T} \!\! d\tau \int_\bx \; \left(
\frac{1}{4}F^a_{\mu\nu}F^a_{\mu\nu}\right.\nonumber\\
&&\left. + \sum_{l=1}^{n_f} \bar{q}_lD\!\!\!\!/\,q_l
\right),
\label{pNRQCD}
\end{eqnarray}
where $\int_\by=\int d^3y$, the trace is over colour indices, $\mathrm{S}=S/\sqrt N$ and $\mathrm{O}=\sqrt2O^aT^a$ are the colour-singlet and colour-octet  quark-antiquark fields and $E^i=F^{i0}$ is the chromoelectric field. The fields S and O depend on the relative distance $\br$, on the center-of-mass coordinate $\bx$ and on the Euclidean time $\tau$. $V_A$, $V_B$ and $V_C$ are the matching coefficients of the first operators in the multipole expansion; for our purposes they are equal to unity. $V_s$ and $V_o$ are the singlet and octet potentials in pNRQCD. The former is known to three loops \cite{Anzai:2009tm}, whereas the latter to two loops \cite{Kniehl:2004rk}. For our purposes the two-loop result \cite{pot2loop} is sufficient. Finally $\delta\mathcal{L}_\mathrm{pNRQCD}$ contains operators in the multipole expansion of order higher or equal to $r^3$, which are shown not to contribute to the Polyakov loop correlator at the intended accuracy.\\
We are then able to write the connected Polyakov loop correlator in terms of singlet and octet correlators as
 \begin{eqnarray}
C_\mathrm{PL}(r,T)
&=& 
\frac{1}{\nc^2}\Bigg[
Z_s \langle S(\br,{\bf 0},1/T)S^\dagger(\br,{\bf 0},0)\rangle
\nonumber\\
&&  +
Z_o \langle O^a(\br,{\bf 0},1/T)O^{a \, \dagger}(\br,{\bf 0},0)\rangle
\nonumber\\
&& 
+ {\cal O}\left(\als^3(rT)^4\right)\Bigg] - \langle L_F \rangle^2.
\label{PLC-pNRQCD}
\end{eqnarray}
The suppressed terms in the last line originate from the $\delta\mathcal{L}_{pNRQCD}$ terms in the Lagrangian. We have thus shown that at the leading order in the multipole expansion of pNRQCD the Polyakov loop correlator can be expressed as a sum of singlet and octet correlators. The matching at the scale $1/r$ furthermore  yields $Z_s=Z_o=1$, consistently with the spectral decomposition of the Polyakov loop correlator \cite{Jahn:2004qr,Luscher:2002qv}, whereas the $\langle SS^\dagger\rangle$ and $\langle OO^\dagger\rangle$ correlators read
\begin{eqnarray}
\langle S\,S^\dagger\rangle &=& e^{-V_s(r)/T}(1+ \delta_s)\equiv e^{-f_s(r,T,m_D)/T},
\label{SSpNRQCD}
\\
\langle O^a\,O^{a\,\dagger}\rangle &=&
e^{-V_o(r)/T}\left[(\nc^2-1)\, \langle L_A \rangle  +  \delta_o\right]\nonumber\\
&\equiv& e^{-f_o(r,T,m_D)/T},
\label{OOpNRQCD}
\end{eqnarray}
where we have defined the gauge-invariant singlet and octet free energies $f_s$ and $f_o$ from these correlators, $\langle L_A\rangle$ is the Polyakov loop in the adjoint representation and $\delta_s$ and $\delta_o$ stand for the loop corrections to these correlators. $\langle L_F\rangle$ $\langle L_A\rangle$, $\delta_s$ and $\delta_o$ encode thus the contribution of the lower scales, in particular $T$ and $m_D$, that are still dynamical in the EFT. We can proceed to compute $\delta_s$ and $\delta_o$ by computing loop diagrams in pNRQCD, again separating the contribution of $T$ from the one of $m_D$ with  the previously exposed techniques. In the end, plugging back the results for $\delta_s$, $\delta_o$ (see \cite{plc}) and those for the Polyakov loops (see Eq. \eqref{finalg4loop}) in Eq. \eqref{PLC-pNRQCD}, large cancellation happen between those terms and we reobtain the result for $C_\mathrm{PL}(r)$ previously shown in Eq. \eqref{finalcpltot}, which was obtained from a direct perturbative computation.\\
In comparison with the recent developments in the construction of a real-time finite-temperature pNRQCD framework for the study of in-medium heavy quarkonium \cite{Brambilla:2008cx,Brambilla:2010vq}, we observe that the free energies $f_s$ and $f_o$ do not agree completely with the real parts of the real-time potentials therein computed. The difference can be traced back to the different boundary conditions: cyclic imaginary time in this case, large real time in that case.\\
EFT-based calculations of the correlator were performed in the past for different scale hierarchies; beside the already mentioned work of Nadkarni for $1/r\sim m_D$ in EQCD \cite{Nadkarni:1986cz}, we also have a calculation within MQCD in the magnetic screening region $m_D\gg 1/r$ \cite{Braaten:1994qx}. In both cases the scale $1/r$ was however not integrated out and the complexity of the bound-state dynamics remained implicit in the correlator.\\
Finally  in Ref. \cite{Jahn:2004qr} the gauge structure and transformation properties of allowed intermediate states contributing to the correlator of Polyakov loops were analyzed; we observe that our results are in agreement with this analysis, since in the weak-coupling regime we considered, both our colour singlet and colour octets transform according to the allowed transformation; for strongly-coupled pNRQCD the allowed intermediate states would be the colour singlet and its hybrid gluonic  excitations.

\emph{Acknowledgements:} We acknowledge financial support from the RTN Flavianet MRTN-CT-2006-035482 (EU) and from the DFG cluster of excellence ``Origin and structure of the universe'' (\href{http://www.universe-cluster.de}{www.universe-cluster.de}).

\end{document}